\documentstyle[psfig,conf_iap,10pt]{article}

\begin{document}

\heading{Exploring the Damped Lyman-$\alpha$ Clouds with AXAF}

\par\medskip\noindent

\author{Taotao Fang, Claude R. Canizares}

\address{Center for Space Research, Massachusetts Institute of
  Technology, 77 Massachusetts Avenue, Cambridge, MA 02139-4307,
  U.S.A}

\begin{abstract}
The High Energy Transmission Grating (HETG) Spectrometer on the
Advanced X-ray Astrophysics Facility (AXAF) (scheduled for launch in
August, 1998) will provide a new tool for the study of absorption in
the X-ray spectra of high redshift quasars due to the material along
the line of sight. In this paper we try to explore the possibility of
using AXAF HETG to detect resonance absorption lines from the Damped
Lyman-$\alpha$ (DLA) clouds. 
\end{abstract}

\section{The simulated spectrum}
AXAF HETG \footnote{See web page
http://asc.harvard.edu/} is
designed to provide high
resolution spectroscopy up to E/$\Delta$E$\sim1000$ (for point source)
between 0.4 keV and 10 keV. 
If the DLA systems contain
a sufficient column density of highly ionized metals, the high
spectral resolution of the HETG will permit detection of resonance
lines. For specificity, we consider Q2223-052, a high redshift quasar at z =
1.404 which has a DLA cloud at $z=0.484$. X-ray
observations with $Einstein$ \cite{wilkes} show its flux to be 
$9.5\times10^{-12} ergs \ s^{-1}cm^{-2}$ between 0.16-3.5 keV. This
implies a count rate 
of 0.36 counts/sec with AXAF MEG in the first order.
Given the observation time and the instrument resolving power we can
calculate the
minimum detectable equivalent width of an absorption feature and the
required ion column density. In Table 1
$N_{1}$ is the required column density. Here $\tau$ is the
optical depth at the line center, and the velocity dispersion $b=200
km/sec$. All energies are in the observer frame. We assume the spectrum of the
high-z quasar has the form of a power law + Galactic absorption + an
assumed 
resonance line from the DLA system due to Si~{\sc xiii} at $1.26 keV$
in the observer frame, then fit it with
a model not
containing the absorption line. In the $\chi^{2}$ plot we
can clearly find the line. \\

\begin{figure}
\centerline{\vbox{
\psfig{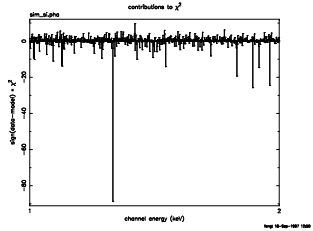}
}}
\caption[]{Q2223-052, the residual $\chi^{2}$ plot}
\end{figure}

\begin{center}

\begin{tabular}{|l|l|l|l|l|l|l|l|}
  \multicolumn{6}{c}{{\bf Table 1.} Two Resonance Lines in AXAF Band}
  \\ \hline
  ion&E(keV)&EW(eV)&$\tau$&$N_1(\times 10^{16})$&$N_2(\times 10^{16})$\\ \hline 
  Mg XI&0.91&1.4&3&6&9.8 \\ \hline
  Si XIII&1.26&1.2&1.2&3.2&4.6 \\ \hline
\end{tabular}

\end{center}

\section{Can we detect DLA systems?}
Table 1 gives the minimum column density which could be detected the
by AXAF HETG. Could a DLA cloud gives such a column density? To
answer this question, we have simulated the ionization structure using
the CLOUDY package \cite{f}. We assume the DLA cloud is photonionizated by
a power law continuum radiation with a mean intensity of
$J_{\nu}$ at 1 Rydberg, $N_{HI}=10^{21}cm^{-2}, n_{H}=10^{-4}cm^{-3}$. The
temperature and metallicity are fixed at $T=2\times10^{4} K$ and
$[Z/H]=-0.5$ \cite{ll}. The value of $J_{\nu}$ is very uncertain, especially at low
redshift \footnote{e.g. See Bechtold's contribution to this
proceeding.}. Our calculation shows that only when $J_{\nu} \geq
5\times10^{-20}ergs \ s^{-1}cm^{-2}Hz^{-1}sr^{-1}$ are ion column
densities high enough to produce resonance lines which could be
detected by AXAF HETG (see Fig.1). The results presented in the sixth
column of Table 2 is calculated with $J_{\nu}=5\times10^{-20}$. Haardt \& Madau \cite{hm} give a value of $J_{\nu}=10^{-22}$ at
$z=0.5$ when taking into account QSOs as the source of
ionization. However, some other photonionization sources, such as the
young 
stars inside the  DLA clouds and nearby quasars, may significantly increase
the intensity of ionization photon flux. On the other hand, there may
also exist other sources of ionization (such as collisional
ionization) or other ionized material co-located with DLA clouds,
which would then simply be markers of higher density regions in an
ionized intergalactic medium. \\

\acknowledgements{We are grateful to the MIT/ASC team for their
  assistance 
  with MARX and XSPEC 10.0 and for helpful discussions. This work was
  supported in part by NASA contract, NAS 8.38249.}

\vfill

\end{document}